\RequirePackage{fix-cm}
\documentclass[twocolumn]{svjour3}          % twocolumn
\smartqed  % flush right qed marks, e.g. at end of proof
\usepackage{graphicx}
\usepackage{setspace}
\usepackage{amssymb,mathrsfs}
\usepackage{bm}
\usepackage{cite}
\usepackage{authblk}
\usepackage[round]{natbib}
\usepackage{color}
\graphicspath{{./figures/}}
\begin{document}

\title{Wicking through a confined micropillar array}



\author{Baptiste Darbois Texier  \and Philippe Laurent \and Serguei Stoukatch \and St\'ephane Dorbolo}


\institute{B. Darbois Texier \and S. Dorbolo \at
              Grasp, Physics Dept., ULg, Li\`{e}ge , Belgium \\
              \email{bdarbois@ulg.ac.be}           %  \\
           \and
           P. Laurent \and S. Stoukatch \at
           Microsys, Montefiore Institute, ULg, Li\`{e}ge , Belgium
}

\date{Received: date / Accepted: date}

\maketitle

\begin{abstract}
This study considers the spreading of a Newtonian and perfectly wetting liquid in a square array of cylindric micropillars confined between two plates. We show experimentally that the dynamics of the contact line follows a Washburn-like law which depends on the characteristics of the micropillar array (height, diameter and pitch). The presence of pillars can either enhanced or slow down the motion of the contact line. A theoretical model based on capillary and viscous forces has been developed in order to rationalize our observations. Finally, the impact of pillars on the volumic flow rate of liquid which is pumped in the microchannel is inspected.

\end{abstract}

\section*{Introduction}

Wicking can be defined as the spreading of a liquid in the tiny structures of a porous media due to capillary forces. The pioneer work concerning the spreading of a viscous liquid in a capillary tube is attributed to \cite{washburn1921dynamics}. The wicking phenomenon has a large spectra of applications such as textile science (\cite{kissa1996wetting}) or heat pipe design (\cite{tien1971minimum}). The problem was revisited because of microelectronics components bonding (\cite{schwiebert1996underfill}). One way to model the wicking process is to consider the spreading of a liquid into a micropillar array. \cite{ishino2007wicking} studied experimentally how a wetting liquid propagates along solid surfaces decorated with a forest of micropillars. This problem leads to numerous experimental derivations (\cite{xiao2011microscale,kim2011liquid,mai2012dynamics,spruijt2015liquid}), numerical contributions (\cite{semprebon2014pinning}) and theoretical modelizations (\cite{hale2014optimization,hale2014capillary}). In the case of microchannels, studies have inspected the effect of surface patterned by posts or pillars on the dynamics of liquid inside the microchannel (\cite{gamrat2008experimental,mognetti2009capillary,mohammadi2013pressure}). Besides, the liquid spreading into a confined micropillar array is involved in various applications such as lab-on-a-chip chromatography (\cite{op2012impact}), isolation of tumor cells (\cite{nagrath2007isolation}) or flow regulation in microfluidic (\cite{vangelooven2010design}). If the problem of liquid impregnation in a micropillar array with a free surface has been extensively studied \cite{bocquet2007flow}, the effect of confinement on this phenomenon is still an open question that we address in this article. 

In this work we consider the case of a square array of cylindric micropillars confined between two plates. We study experimentally the spreading of a perfectly wetting liquid as a function of the pillars characteristics (height, pitch, diameter). The measurements are compared with a theoretical model based on capillary forces and viscous resistance. Finally, the consequences of this work for capillary pumping is discussed. In addition, more complex micropillars arrangements are examined such as non-square and non-uniform lattices.

\section{Experiments}\label{sec:experiments}

\begin{figure}[htp!]
\centering
\includegraphics[width=7cm]{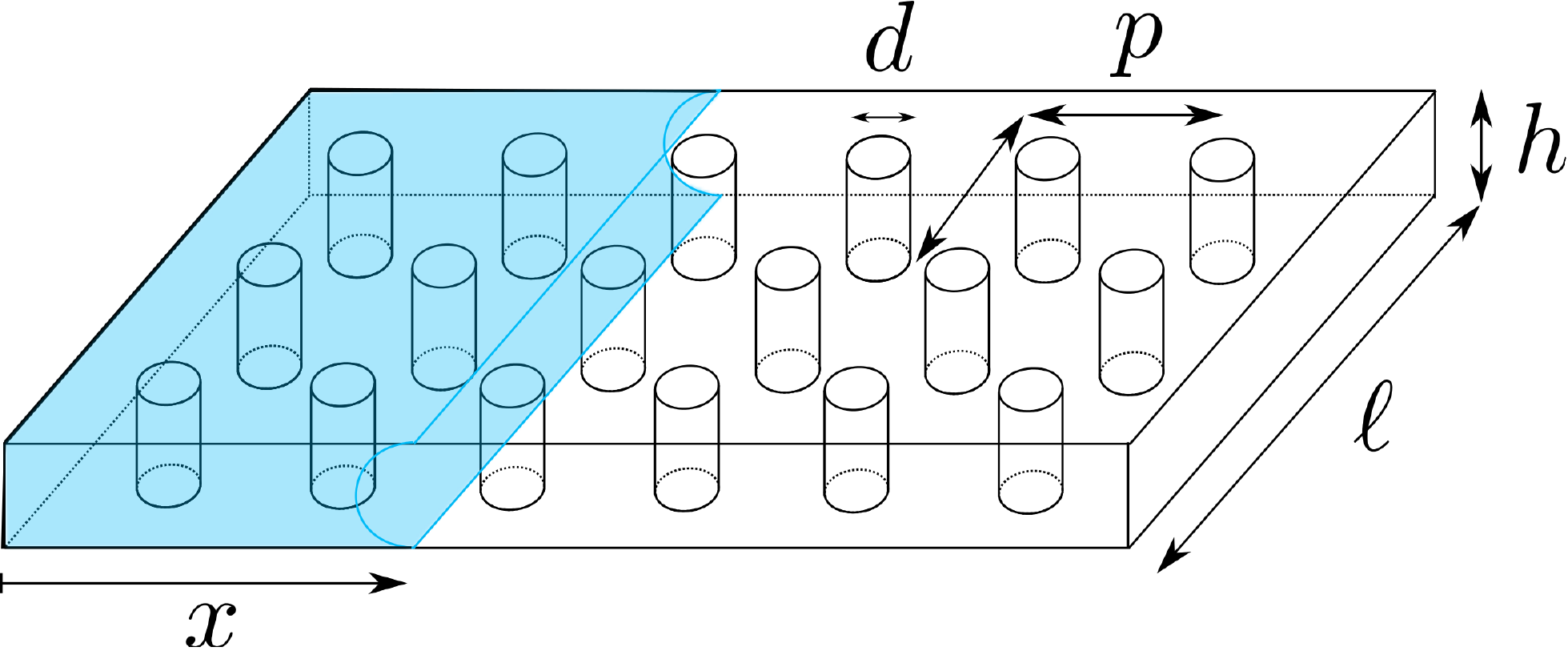}
\caption{Sketch of the experimental setup. The channel height is $h$, its width is $\ell$. The channel contains an array of pillars of diameter $d$ spaced by a pitch $p$. The height of the pillars is the same that the one of the channel. The dynamic viscosity of the fluid which spreads in the channel is denoted $\eta$. The mean position of the contact line is $x$.}
\label{fig:sketch}
\end{figure}

\begin{figure}[h!]
\centering
		(a) \includegraphics[width=7cm]{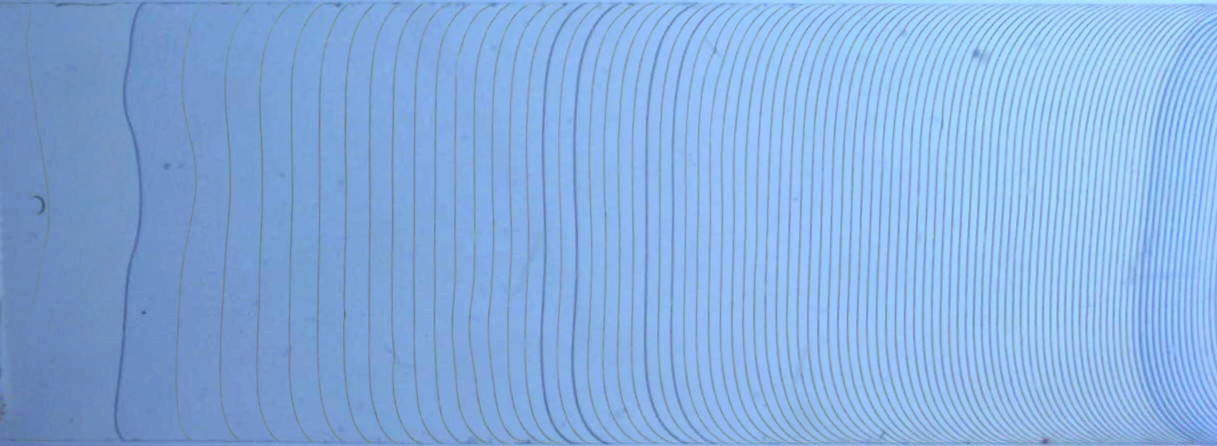}\\
		(b) \includegraphics[width=7cm]{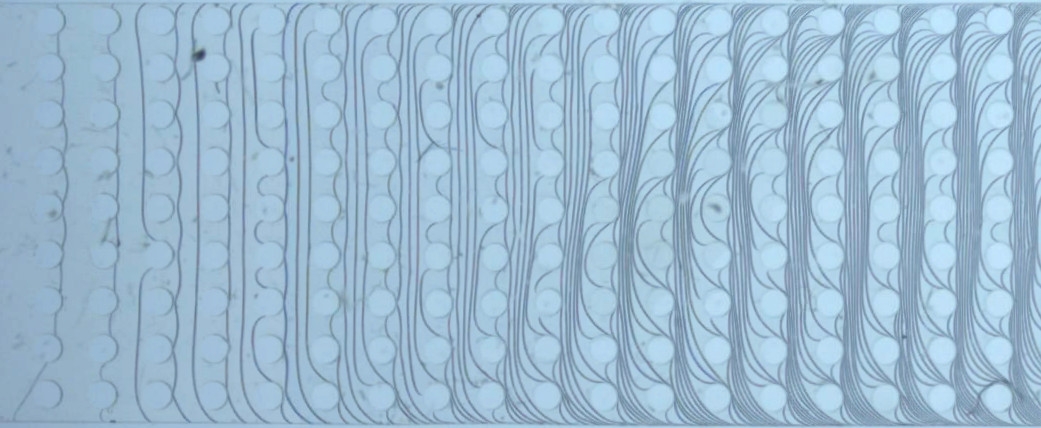}\\
 \caption{Superimposition of top-view pictures of liquid impregnation of microchannels over time. The time step between two frames is 1 s. The fluid is Newtonian silicone oil V100. (a) Empty microchannel of height $h=80 \, \rm{\mu m}$. (b) Microchannel of height $h=80 \, \rm{\mu m}$ filled with pillars of diameter $d=400  \, \rm{\mu m}$ separated by a pitch $p=  800\, \rm{\mu m}$.}
\label{fig:chrono}
\end{figure}

\begin{figure}[h!]
\centering
		\includegraphics[width=7.5cm]{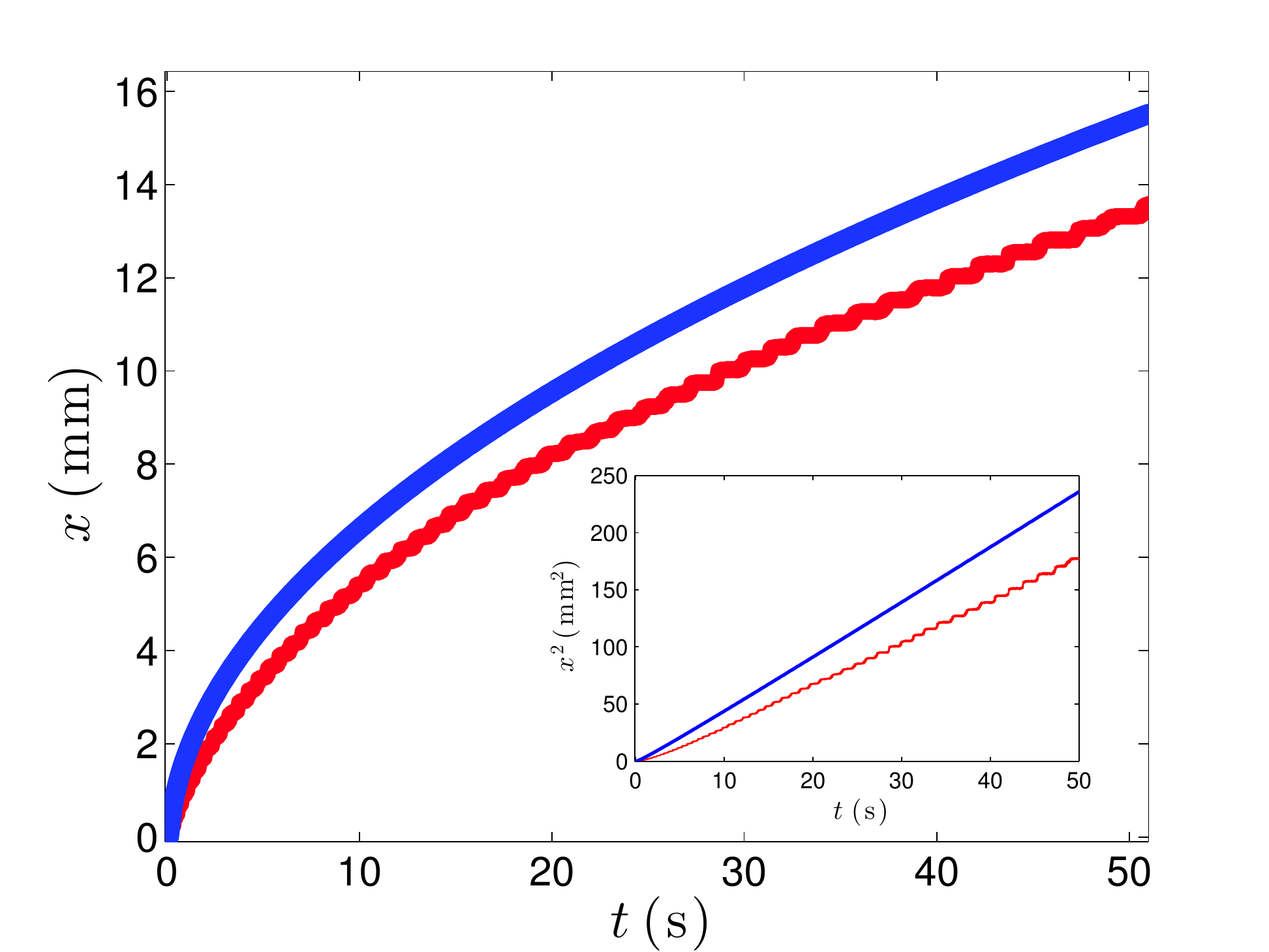}
 	\caption{Mean position of the contact line $x$ as a function of time $t$ for a silicone oil V100 ($\rho=980 \, \rm{kg/m^3}$, $\gamma=23  \, \rm{mN/m}$ and $\eta=100 \, \rm{mPa \cdot s}$) which spreads in two different microchannels. Each channel has the same height $h=80 \, \rm{\mu m}$ but a different density of pillars. The blue dashed line shows the case of an empty microchannel and the red solid line stands for a microchannel with pillars of diameter $d=400  \, \rm{\mu m}$ separated by a pitch $p=  800\, \rm{\mu m}$. The inset shows the square of the mean contact line position $x^2$ as a function of time for the two previous experiments.}
		\label{fig:t_x}
\end{figure}

Microchannels were obtained by negative SU-8 photolithography made on a silicon wafer (\cite{del20078}) prior to a PDMS molding (\cite{folch1999molding}). The PDMS cavity was bonded on a substrate of the same material by oxygen plasma exposure. By varying the design of a the mask used in the photolithography, we changed the geometry of the micropillars array inside the channel [Fig. \ref{fig:sketch}]. The channel height $h$ could be changed between $50 \, \rm{\mu m}$ and $100 \, \rm{\mu m}$, the pillar diameter $d$ from $80 \, \rm{\mu m}$ to $1000 \, \rm{\mu m}$ and the lattice pitch $p$ between $150 \, \rm{\mu m}$ and $1000 \, \rm{\mu m}$. Thus, the pillar aspect ratio $h/d$ may vary between 0.05 and 1.2 and the ratio $p/d$ between 1.5 and 10. The transverse dimension of the channel $\ell$ was chosen in order to keep the ratio $\ell/h$ larger than 100. These PDMS microchannels were put into contact with a liquid reservoir containing a Newtonian silicone oil V100 of density $\rho=980 \, \rm{kg/m^3}$,  dynamic viscosity $\eta=100 \, \rm{mPa \cdot s}$ and surface tension $\gamma \simeq 23 \pm  0.3 \, \rm{mN/m}$ at a temperature $T=20 ^\circ \rm{C}$. The mean position of the contact line was recorded from above and reported in Fig. \ref{fig:chrono}. The analysis of these experiments provides the position of the contact line $x$ over time as shown in Fig. \ref{fig:t_x}. 

One observes that, whatever the considered microchannel, the mean position of the contact line propagates according to the square root of time. This dependency is underlined by the linearity of $x^2$ with time as shown in the inset of Fig. \ref{fig:t_x} and corresponds to a Washburn-like law expected for the spreading of a viscous fluid induced by capillary forces (\cite{washburn1921dynamics}). Besides, one notices the influence of the pillars array on the dynamics of the contact line by a change of the pre-factor of the square root law. In the considered case, the presence of pillars in the channel slows down the liquid dynamics. Additionally, the introduction of pillars in the microchannel induces a stick-slip behavior in the dynamics of the contact line which appears by the way of steps in the red curves in Fig. \ref{fig:t_x}.

\begin{figure}[h!]
\centering
	\begin{minipage}[c]{0.5\textwidth}
  		\centering
  		\hspace{0.5cm}(a)\\
		\vspace{0cm}
		\includegraphics[width=7cm]{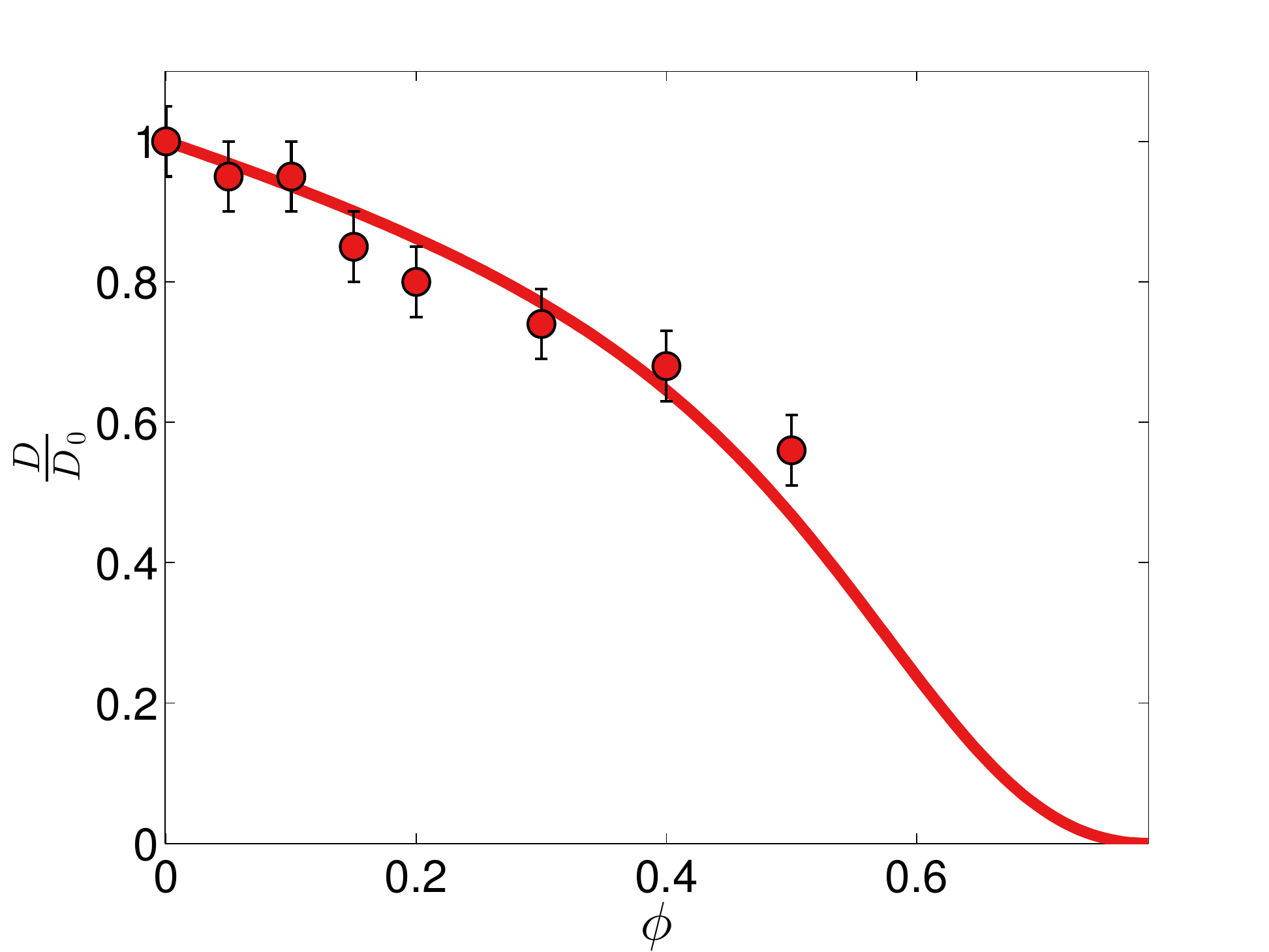}
 	\end{minipage}\\
	\hspace{1cm}
	\begin{minipage}[c]{0.5\textwidth}
  		\centering
  		\hspace{0.5cm}(b)\\
		\vspace{0cm}
		\includegraphics[width=7cm]{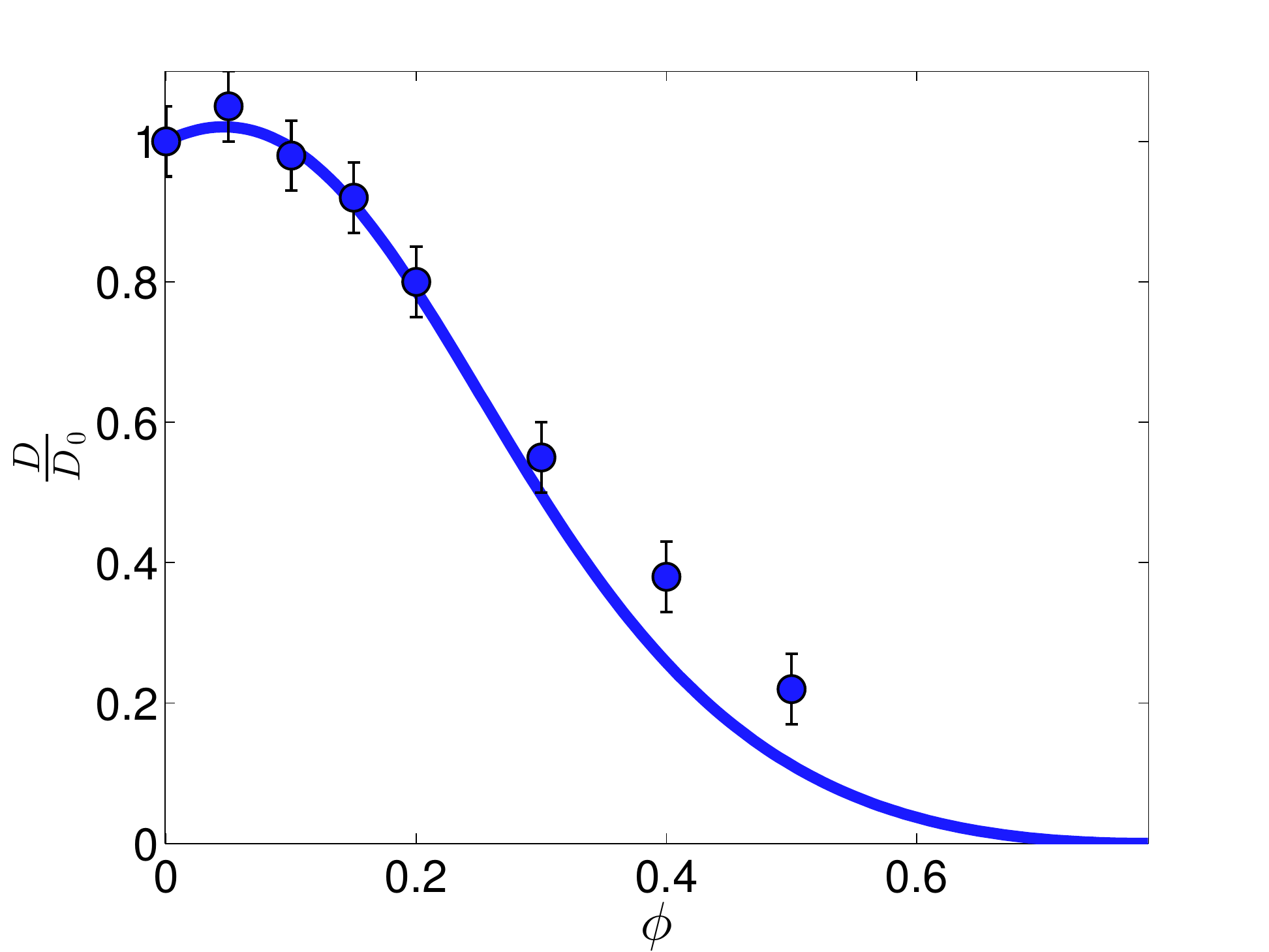}
	\end{minipage}%
 \caption{Diffusivity ratio $D/D_0$ as a function of the pillars density  $\phi=\pi d^2 / 4 p^2$ for two different pillar aspect ratios $h/d$. Dots correspond to the flow of silicone oil V100 ($\rho=980 \, \rm{kg/m^3}$, $\gamma=23  \, \rm{mN/m}$ and $\eta=100 \, \rm{mPa \cdot s}$) into micro-channels of height $h=80 \, \rm{\mu m}$ with pillars of diameter $d=400  \, \rm{\mu m}$ ($h/d=0.2$) in figure (a) and pillars of diameter $d=80\,\rm{\mu m}$ ($h/d=1$) in figure (b). Solid lines correspond to the theoretical model developed further in Section \ref{sec:model} and show the solution of equation (\ref{eq:ratio}) for different pillars aspect ratio ($h/d=0.2$ for the red solid line in figure (a) and $h/d=1$ for the blue solid line in figure (b)).} 
\label{fig:permeability_ratio}
\end{figure}

Thereafter, the effect of the micropillar array on the spreading dynamics is quantified. We measure the pre-factor $D$ of the square root law $x=\sqrt{Dt}$ for microchannels as a function of the lattice properties (pillar diameter $d$, height $h$ and pitch $p$). This pre-factor defines the diffusivity of the liquid in the micro-channel and is measured by fitting the slope of $x^2 \, (t)$ by the way of a least mean squares method. The analogous diffusivity of the liquid in the empty microchannel is denotes $D_0$. Symbols in Fig. \ref{fig:permeability_ratio}a and \ref{fig:permeability_ratio}b show the evolution of the experimental ratio $D/D_0$ as a function of the pillar density $\phi=\pi d^2 / 4 p^2$ where $d$ and $h$ are kept constants and only $p$ is varied. Two sets of experiments have been done for a different pillar aspect ratio of $h/d=0.2$ and $h/d=1$ and are plotted with dots.

As expected, when the pillar density tends to zero, the fluid diffusivity $D$ recovers the one of an empty microchannel $D_0$. In the case for which the pillars aspect ratio $h/d=0.2$, the diffusivities ratio $D/D_0$ decreases monotonically with the pillar density $\phi$. The situation is more complicated for the case of $h/d=1$ because the ratio $D/D_0$ first increases up to pillar density of about 0.06 and then decreases to reach values lower than unity for large pillar densities ($\phi > 0.15$). One remarks that in the situation where $h/d=1$ and $\phi<0.15$, the presence of pillars enhances the spreading of the liquid in the channel.

\section{Model}\label{sec:model}

\subsection{Empty cavity}

The goal of this section is to develop a theoretical model in order to rationalize the experimental observations. First, we evaluate the dimensionless numbers associated to previous experiments in order to determine the main forces in presence. According to experimental data presented in Section \ref{sec:experiments}, the spreading velocity reaches $\dot{x} \sim 1 \, \rm{mm/s}$ in less than 20 ms. This allows to estimate the Reynolds number associated to the fluid flow at this particular time: $Re= {\rho h \dot{x} }/{\eta} \sim {10^3 \times 10^{-4} \times 10^{-3}}/{0.1} \sim 10^{-3}$. Therefore, the liquid flow is dominated by viscous friction and no inertial terms will be taken into account in this study. The experimental Bond number, which compares the relative effect of gravity and surface tension, is $Bo={\rho g h^2}/{\gamma} \sim {10^3 \times 10 \times 10^{-8}}/{10^{-2}} \sim 10^{-2}$. Also, we estimate the capillary number $Ca=\eta U /\gamma \simeq 0.1 \times 10^{-3} / 2 \times 10^{-2} \simeq 5 \times 10^{-3}$. The small value of the capillary number compared to one signifies that the contact angle remains close to its equilibrium value (\cite{thompson1989simulations}). Finally, the experimental data presented in Section \ref{sec:experiments} will be approached by a model which balances capillary and viscous forces. 

\begin{figure}[h!]
\centering
	\begin{minipage}[c]{0.45\textwidth}
  		\centering
  		\hspace{0.5cm}(a)\\
		\vspace{0.2cm}
		\includegraphics[width=6.5cm]{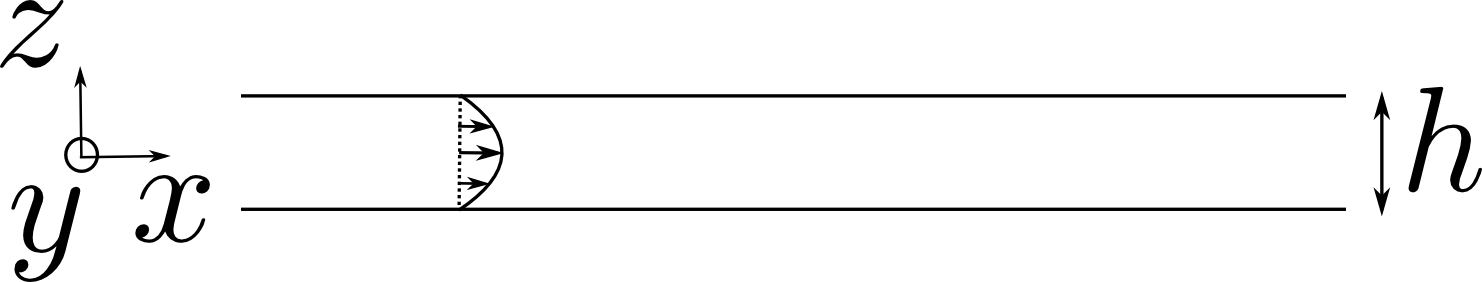}
		\vspace{1cm}
 	\end{minipage}\\
	\begin{minipage}[c]{0.45\textwidth}
  		\centering
  		(b)\\
		\vspace{0.3cm}
		\includegraphics[width=4cm]{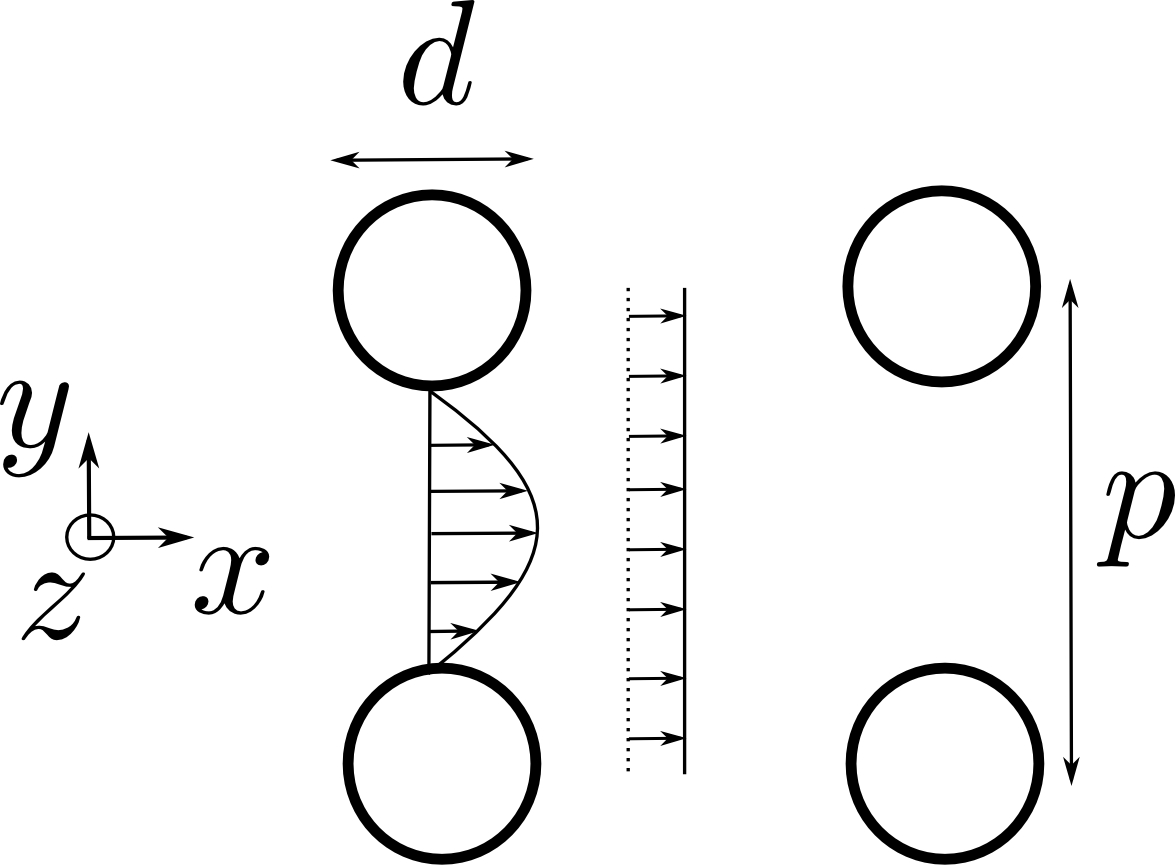}
 	\end{minipage}
 \caption{(a) Sketch of the liquid flow in an empty microchannel viewed from the side. (b) Sketch of the assumed liquid flow in-between four micropillars viewed from above.}
\label{fig:sketch_model}
\end{figure}

In the case where the microchannel is empty, the surface wetted by the fluid at a positon $x$ is equal to $S=2xl$ (neglecting the channel height $h$ which is more than 100 times smaller than its width $\ell$). The resulting capillary force in the $x$ direction is $f_\gamma =\gamma \partial S / \partial x=2 \gamma l$. The non-slipping boundary conditions of the fluid on the two microchannel plates impose a Poiseuille flow which can be expressed as: $v_x = v_0 \left( 1 - 4 z^2/h^2 \right)$ [Fig. \ref{fig:sketch_model}a]. The flow conservation imposes $v_0=3 \dot{x}/2$. The viscous force along the $x$-direction is equal to $f_x= \eta \int_{x'=0} ^{x'=x} \int_{y'=-l/2} ^{y'=l/2} \int_{z'=-h/2} ^{z'=h/2} \Delta v_x \, dx' dy' dz' $ which provides $f_x= -12 \, \eta {x \dot{x} l}/{h}$. Balancing the capillary force with the viscous contribution provides

\begin{equation}
x^2 (t) = \frac{\gamma h}{3 \eta} \, t
\label{eq:empty1}
\end{equation}

This approach predicts the square root evolution of the contact line position with time observed experimentally in Section \ref{sec:experiments} and provides a theoretical value of the diffusivity $D_0=\gamma h / 3 \eta$ in the case of an empty microchannel. Note that, in practice this law is used to estimate the flowing time of the underfill process in microelectronics (\cite{wan2008experimental}).

\subsection{Effect of the pillar array}\label{sec:model_pillars}

When the pillars are present in the microchannel, the surface of the cavity wetted by the fluid reads

\begin{equation}
S=2xl- 2 \pi \left( \frac{d}{2} \right)^2 \frac{x}{p} \frac{l}{p} + \pi d h \frac{x}{p} \frac{l}{p}
\label{eq:S}
\end{equation}

\noindent where $x/p$ is the mean number of pillar rows wetted by the fluid when its interface is located in $x$ and $l/p$ the mean number of wetted pillar rows along the transverse direction of the microchannel. The global approach used here is justified because we consider the mean dynamics of the contact line over a distance $x$ much larger than the characteristics of the pillar array ($p$ and $d$).

The resulting capillary force is

\begin{equation}
f_\gamma=2 \gamma l \left[ 1- \frac{\pi d^2}{4 p^2} +  \frac{h}{d} \frac{\pi d^2}{2p^2} \right]
\label{eq:f_gamma_1}
\end{equation}

\noindent which reduces to

\begin{equation}
f_\gamma=2 \gamma l \left[ 1+ \left(2 \overline{h} -1 \right) \phi \right]
\label{eq:f_gamma_2}
\end{equation}

\noindent with $\phi=\pi d^2 / 4 p^2$ the pillar density and $\overline{h}=h/d$ the pillar aspect ratio.

The presence of pillars inside the microchannel modifies the fluid velocity profile inside the cavity and thus the viscous force experienced by the fluid. Various studies inspected experimentally and theoretically the pressure drop resulting of a liquid flow through a square arrays of cylinders embedded inside a microchannel (\cite{sadiq1995experimental,tamayol2009analytical,gunda2013measurement,tamayol2013low}). However, these studies only consider the situation where the vertical confinement is negligible. In this paper, we estimate the viscous force by assuming a Poisseuille flow in the space between micropillars. We suppose that the fluid velocity profile in-between pillars is $v_{x p}=v_{0 p} \left( 1 - 4 y^2 / (p-d)^2\right)$ $\left( 1 - 4 z^2 / h^2\right)$ whereas the fluid velocity profile elsewhere remains $v_{x}= v_0 \left( 1 - 4 z^2 / h^2\right)$ [Fig. \ref{fig:sketch_model}b]. The flow mass conservation yields $v_{0p}=9 p \dot{x}/4 (p-d)$ and $v_0= 3 \dot{x}/2$. Finally, the viscous force resulting from the fluid flow between two pillars reads

\begin{equation}
f_{x p} = -12 \, \eta \, \dot{x} \, d \, \frac{p}{p-d} \, \left( \frac{h}{p-d} + \frac{p-d}{h}\right) 
\label{eq:f_z_1}
\end{equation}

The total viscous force in-between pillars when the contact line is located in $x$ can be approached by

\begin{equation}
f_{xp \, tot}  = -12 \, \eta \, \dot{x} \, x  \, l \, \frac{d}{p^2} \, \frac{p}{p-d} \, \left( \frac{h}{p-d} + \frac{p-d}{h}\right)
\label{eq:f_z_2}
\end{equation}

The previous equation yields

\begin{equation}
f_{xp \, tot}  = - \frac{12 \, \eta \, \dot{x} \, x  \, l }{h}  \, \left( \frac{\overline{h}^2}{\sqrt{\frac{\phi_m}{\phi}} \, \left(\sqrt{\frac{\phi_m}{\phi}}-1 \right)^2} +  \sqrt{\frac{\phi}{\phi_m}} \right)
\label{eq:f_z_2}
\end{equation}

\noindent with $\phi_m=\pi/4\simeq 0.78$ the maximal pillar density when $p=d$. The viscous contribution resulting from the fluid flow out of inter-pillar areas is

\begin{equation}
f_{x \, tot}= - \frac{12  \, \eta \, \dot{x} \, x \, l }{h}  \left( 1 - \frac{d}{p} \right)
\label{eq:f_y_1}
\end{equation}

This equation reduced to

\begin{equation}
f_{x \, tot}= -\frac{ 12  \, \eta \, \dot{x} \, x \, l }{h}  \left( 1 - \sqrt{\frac{\phi}{\phi_m}} \right)
\label{eq:f_y_1}
\end{equation}

Balancing the capillary and viscous forces, $f_{xp \, tot} + f_{x \, tot} + f_\gamma=0$, gives

\begin{equation}
\begin{array}{l}
12 \, \eta \frac{x \dot{x} l}{h}  \left[  1 + \frac{\overline{h}^2}{\sqrt{\frac{\phi_m}{\phi}} \, \left(\sqrt{\frac{\phi_m}{\phi}}-1 \right)^2}\right] =\\ 2 \gamma l \left[ 1+ \left( 2 \overline{h} -1 \right) \phi \right] 
\end{array}
\label{eq:motion}
\end{equation}

This relation simplifies as

\begin{equation}
\dot{x^2} = \frac{\gamma h}{3\eta} \, \, \frac{ 1+ \left( 2 \overline{h} -1 \right) \phi }{  1 + {\overline{h}^2}/{\sqrt{\frac{\phi_m}{\phi}} \, \left(\sqrt{\frac{\phi_m}{\phi}}-1 \right)^2}  }
\label{eq:motion2}
\end{equation}

In presence of pillars, the liquid still follows a Washburn-like law ($x = \sqrt{D t}$) but the diffusivity now depends on the pillar array characteristics ($\phi$ and $\overline{h}$). We define the ratio of the diffusivity $D$ of the channel in presence of pillars and the diffusivity $D_0={\gamma h}/{3\eta}$ for an empty channel. This ratio is expressed as

\begin{equation}
\frac{D}{D_0} = \frac{ 1+ \left( 2 \overline{h} -1 \right) \phi }{  1 + {\overline{h}^2}/{\sqrt{\frac{\phi_m}{\phi}} \, \left(\sqrt{\frac{\phi_m}{\phi}}-1 \right)^2}  }
\label{eq:ratio}
\end{equation}

\subsection{Comparison with experiments}

Predictions of Eq. (\ref{eq:ratio}) are compared with experimental data in Fig. \ref{fig:permeability_ratio} for different pillars density $\phi$ and various normalized pillar height $\overline{h}$ by the way of solid lines. The good agreement between experiments and theoretical predictions validates the assumptions made for the profile of the flow in the model developed in section \ref{sec:model_pillars}.

\begin{figure}[htp!]
\centering
\includegraphics[width=8cm]{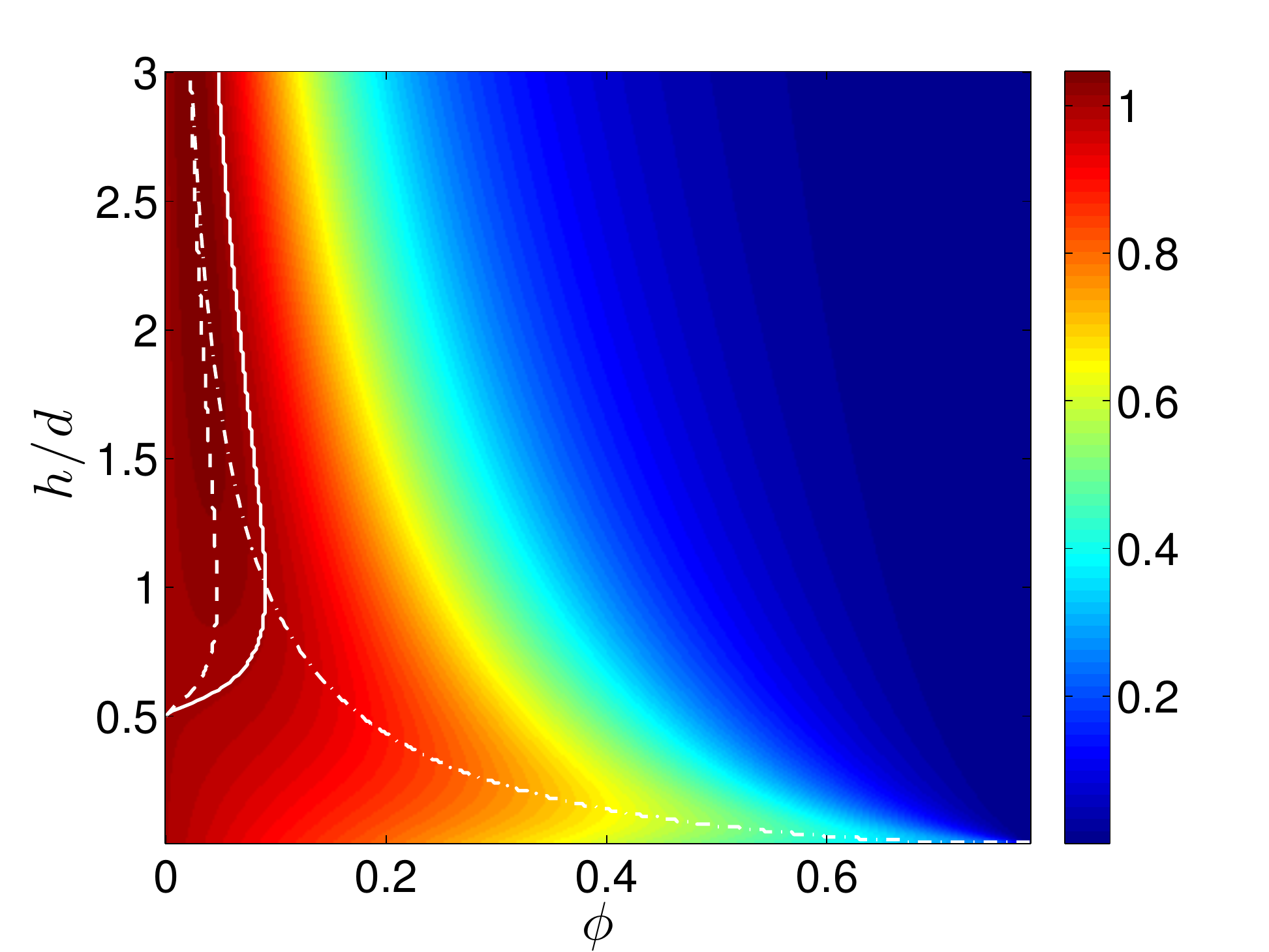}
\caption{Ratio $D/D_0$ as a function of the pillar density $\phi$ and the normalized height of pillars $\overline{h}=h/d$ as predicted by equation (\ref{eq:ratio}). The white solid line separates the region where the diffusivity ratio $D/D_0$ is larger than one. The white dashed line shows the pillar density $\phi_{\rm{max}}$ which maximizes the ratio $D/D_0$ for a given aspect ratio $\overline{h}$ larger than 0.5. The white dotted line indicates the pillar aspect ratio $\overline{h}_{\rm{max}}$ which maximes the ratio $D/D_0$ for a given pillar density. The theoretical predictions of these three lines are detailed in the Appendix A.}
\label{fig:ratio}
\end{figure}

Thereafter, we consider the predictions of the model in a broader range of parameters compared to experiments. Figure \ref{fig:ratio} shows ${D}/{D_0}$ as a function the pillar density $\phi$ and the normalized height $\overline{h}$ as predicted by Eq. (\ref{eq:ratio}). One notices that whatever the pillar aspect ratio, the diffusivity ratio $D/D_0$ tends to one for small pillar densities ($\phi \rightarrow 0$) and falls to zero for large pillar densities ($\phi \rightarrow \phi_m$). If $\overline{h}>0.5$, the diffusivity ratio is slightly larger than one for small density values ($\phi \lesssim 0.15$) as delimited by the white solid line in Fig. \ref{fig:ratio}. The analytical determination of the domain where $D/D_0>1$ is presented in the Appendix A. The maximal value of the diffusivity ratio $D/D_0$ increases with the normalized aspect ratio $\overline{h}$. For $\overline{h}=3$, $D/D_0$ reaches a maximal value of about 1.1 for $\phi = 0.06$. This behavior can be understood by the fact that when the liquid wets a pillar, the gain of surface energy is $\gamma \pi d h$ whereas if there was no pillar the gain of energy would have been $ \gamma \pi d^2 /2$. If $h>d/2$, the presence of pillars increases the energetic benefit of the liquid impregnation and fasten its dynamics.

\section{Discussion}

\subsection{Fluid pumping}

In our experiments, the capillary forces pump the fluid inside the microchannel. The liquid flow rate is defined as $Q={dV}/{dt}$ with $V$ the volume of fluid inside the microchannel. When the contact line has the position $x$, the volume of fluid inside the cavity is $V=lhx-\pi \left( \frac{d}{2} \right)^2 h \frac{l}{p} \frac{x}{p}$. Substituting the relation $x=\sqrt{Dt}$ in the previous definition leads to

\begin{equation}
Q= \frac{h \, l \, (1-\phi)}{2}  \sqrt{\frac{D}{t}}
\label{eq:flowrate}
\end{equation}

The ratio of the flow rate $Q$ in a cavity with pillars and the flow rate $Q_0$ in an empty cavity is

\begin{equation}
\frac{Q}{Q_0}= \left( 1 - \phi \right) \sqrt{\frac{D}{D_0}}
\label{eq:flowrate}
\end{equation}

Inserting equation (\ref{eq:ratio}) in the previous ratio yields

\begin{equation}
\frac{Q}{Q_0}= \frac{ \left( 1 - \phi \right) \left( 1+ \left(2 \overline{h} -1 \right) \phi \right)^{1/2}}{  \left( 1 + {\overline{h}^2}/{\sqrt{\frac{\phi_m}{\phi}} \, \left(\sqrt{\frac{\phi_m}{\phi}}-1 \right)^2}   \right)^{1/2}}
\label{eq:flowrate}
\end{equation}

\begin{figure}[htp!]
\centering
\includegraphics[width=8cm]{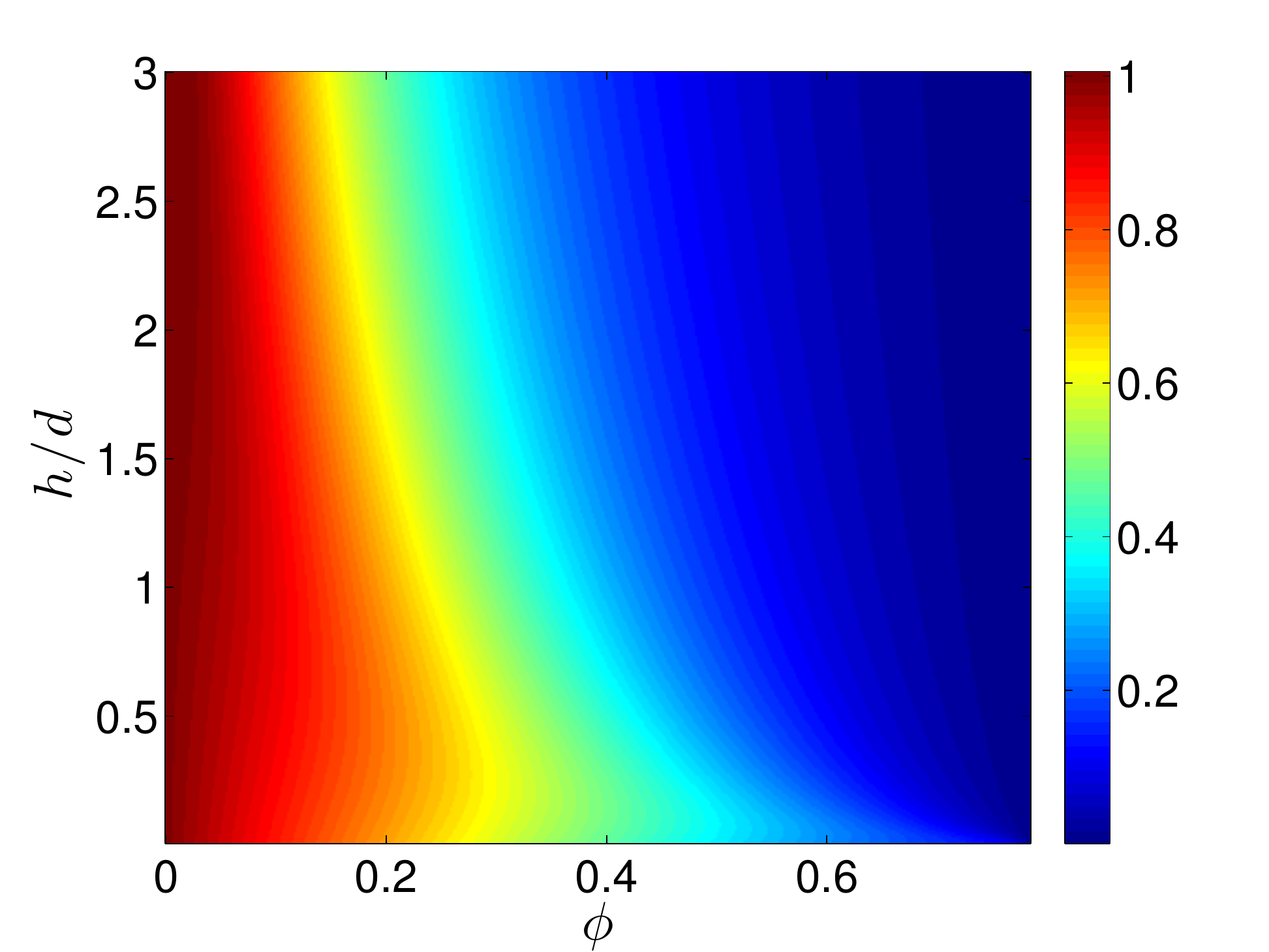}
\caption{Ratio $Q/Q_0$ as a function of the pillar density $\phi$ and the normalized height of the channel $\overline{h}=h/d$.}
\label{fig:ratio_QQ0}
\end{figure}

Figure \ref{fig:ratio_QQ0} presents the flow rate ratio $Q/Q_0$ as a function of the pillar density $\phi$ and the pillar aspect ratio $\overline{h}$. Whatever the ratio $h/d$, if the pillar density is small ($\phi \rightarrow 0$), the flow rate ratio $Q/Q_0$ tends to one. In the opposite situation where the pillar density is large ($\phi \rightarrow \phi_m$), the ratio $Q/Q_0$ falls to zero. One notices that the presence of pillars only slows down the flow rate $Q$ in a cavity with pillars compared to the case of an empty cavity. Thus, the addition of micro-structures in a microchannel does not enhance its pumping properties. This conclusion is of first importance for the development of capillary pumps in the context of autonomous microfluidic. In such a situation, the presence of micro-structures in the capillary pump increases its efficiency by inducing large interface curvatures which produce a large overpressure and drive the fluid (\cite{zimmermann2007capillary}). According to our conclusions, this effect works only if the hydraulic resistance of the system is imposed by an element placed before the pump. In the context of autonomous microfluidic, the determination of the hydraulic resistance of the capillary pump compared to the one of upstream elements is essential to maximize the fluid flow rate. 

\subsection{Other lattices}

\begin{figure*}[htp!]
\centering
	\begin{minipage}[c]{0.28\textwidth}
  		\centering
  		(a)\\
		\vspace{0.2cm}
		\includegraphics[height=3.5cm]{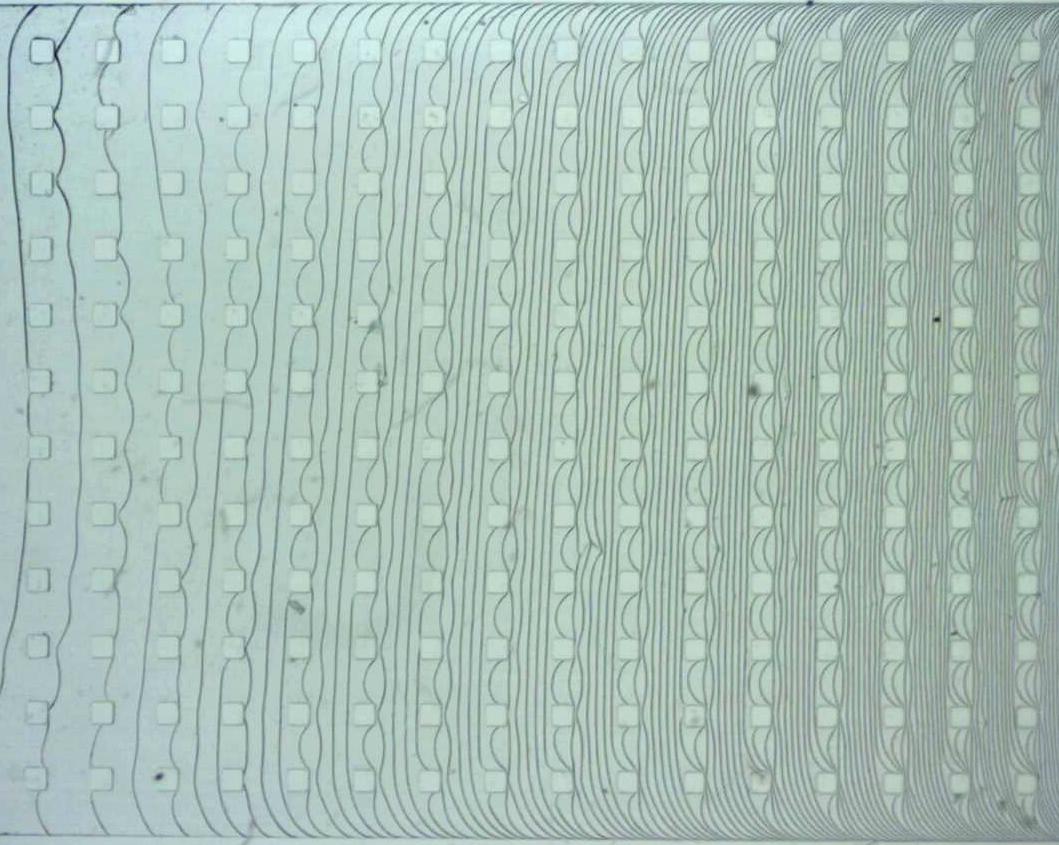}
 	\end{minipage}
	\begin{minipage}[c]{0.25\textwidth}
  		\centering
  		(b)\\
		\vspace{0.2cm}
		\includegraphics[height=3.5cm]{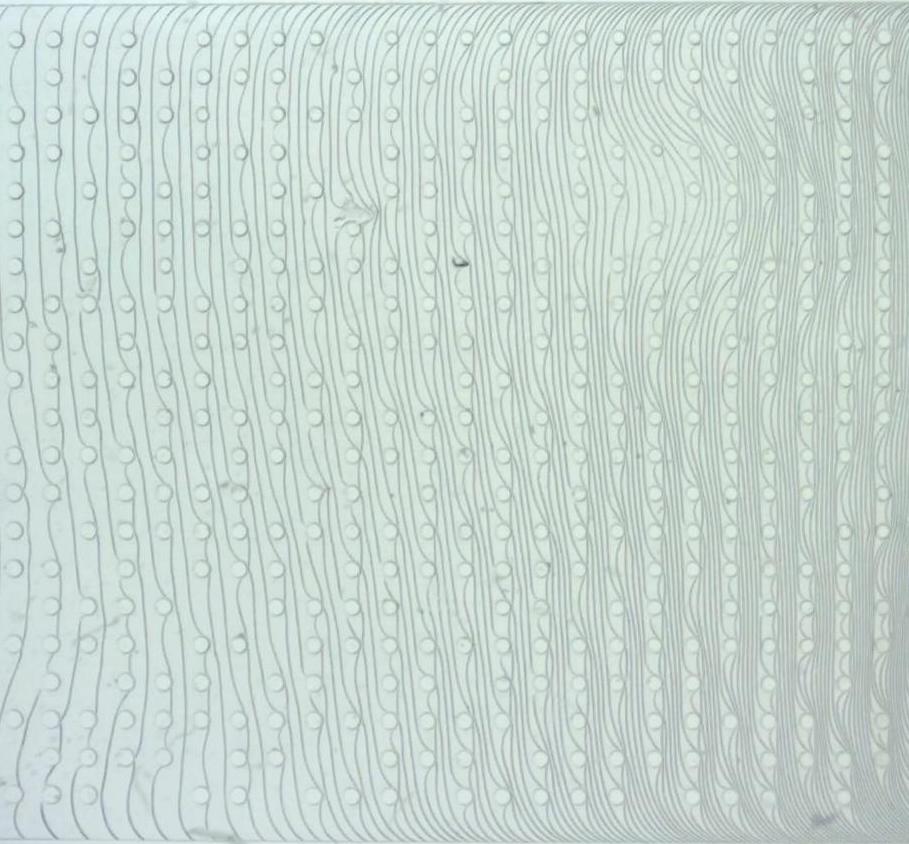}
		%\vspace{1.1cm}
 	\end{minipage}
	\begin{minipage}[c]{0.3\textwidth}
  		\centering
  		(c)\\
		\vspace{0.2cm}
		\includegraphics[height=3.5cm]{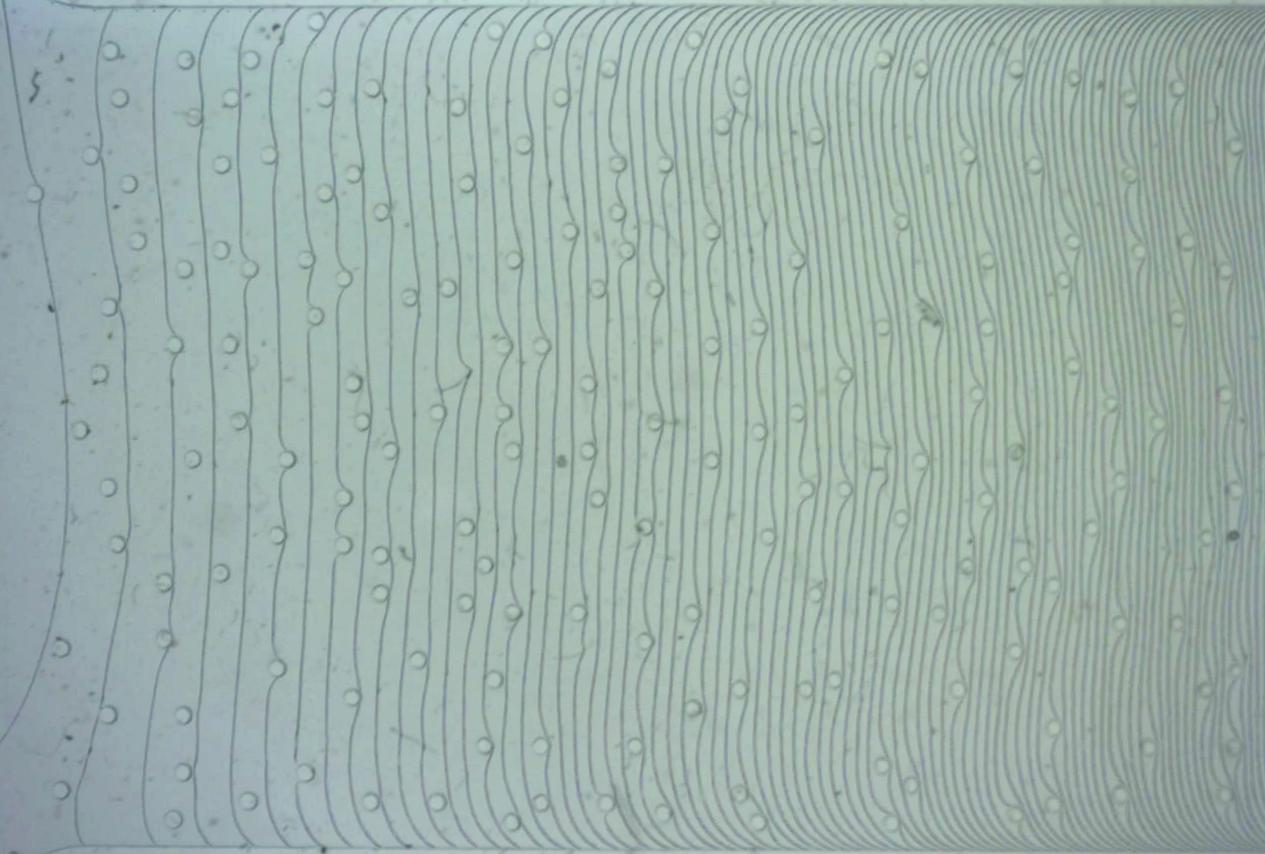}
 	\end{minipage}
 \caption{Examples of the progression of a Newtonian silicone oil of dynamic viscosity $\eta=100 \, \rm{mPa \cdot s}$ in original lattices of micropillars: (a) Square lattice of square pillars of width $d=250 \, \rm{\mu m}$. The pitch between pillars is $p=750 \, \rm{\mu m}$. (b) Square lattice of cylindric micropillars of diameter $d=250 \, \rm{\mu m}$ with defects. The pitch of the lattice is $p=600 \, \rm{\mu m}$. (b) Lattice of randomly distributed cylindric micropillars of diameter $d=250 \, \rm{\mu m}$. In these three experiments, the time step between two positions of the contact line is 1 s.}
\label{fig:exotic lattice}
\end{figure*}

This section aims to discuss how the conclusions drawn previously for square lattices of cylindrical pillars depend on the pillars geometry and arrangement. With the procedure described in Section \ref{sec:experiments} for microchannels fabrication, we create original pillars arrangements such as square lattices of square pillars [Fig. \ref{fig:exotic lattice}a], square lattices of cylindric pillars with defects [Fig. \ref{fig:exotic lattice}b] and randomly distributed lattices of cylindric micropillars [Fig. \ref{fig:exotic lattice}c]. 

For all these cases, we recorded the wicking of a Newtonian silicone oil V100 inside the microchannels. The diffusivity coefficient $D$ was measured and compared with the one of a reference case that we choose to be a square lattice of cylindric micropillars with the same pillar density. For lattices that were non-square with cylindrical pillars, the density of pillars is defined by the ratio between the cross section of all the pillars and the total section of the microchannel. The results of these experiments are listed in Table 1.

\begin{table*}[htp!]
     \label{tab:exotic_lattice}
\begin{center}
    \begin{tabular}{|p{3cm}|p{1cm}|p{1cm}|p{1cm}|p{1cm}|p{1.7cm}|p{1.7cm}|p{1.7cm}|}
    \hline
    Lattice & $d (\rm{\mu m})$ & $p (\rm{\mu m})$ & $\phi$ & $h (\rm{\mu m})$ & $D \, (\rm{m^2/s})$ & $D_{\rm{ref}} \, (\rm{m^2/s})$ & $D/D_{\rm{ref}}$ \\
    \hline
    square section pillars & 250 & 750 & 0.11 & 80 & $5.82 \times 10^{-6}$ & $ 5.94 \times 10^{-6}$ & 0.98 \\
    \hline
    square lattice of cylindric pillars with defects & 250 & 600 &0.12 & 80 &$6.06 \times 10^{-6}$ & $5.88 \times 10^{-6}$ & 1.03 \\
    \hline
    random lattice of cylindric pillars & 250 & -  & 0.05& 80 & $6.2 \times 10^{-6}$ &   $6.07 \times 10^{-6}$ & 1.02 \\
    \hline
    \end{tabular}
     \caption{Measurements of diffusivity coefficients $D$ for a Newtonian silicone oil V100 ($\rho=980 \, \rm{kg/m^3}$, $\gamma=23  \, \rm{mN/m}$ and $\eta=100 \, \rm{mPa \cdot s}$) in original lattices (with square section pillars, square lattice with defects and random lattice of cylindrical pillars). The diffusivity coefficient is compared with a reference case defined as a square lattice of cylindrical pillars with the same density of pillars.}
\end{center}
\end{table*}

First, one notices that in the three situations considered previously, the diffusivity modification induced by a change of pillar geometry or lattice arrangement are moderate. Indeed, the relative variation of the diffusivity stays below 3 \% in the three situation considered above. Thus, the dynamics of the contact line in a microchannel is mainly ruled by the micro-structure density $\phi$ and their aspect ratio $h/d$. In details, we observe that the change in the geometry of the pillars from cylindrical to square section reduces the dynamics of the liquid impregnation. The presence of defects in a square lattice of cylindric micropillars enhances slightly the fluid wicking. Finally, a random distribution of cylindric micropillars is a bit more efficient relatively to a square arrangement. Obviously, a systematic study as a function of the lattice parameters has to be lead in order to conclude on the precise impact of square section pillars, presence of defect or random distribution of pillars.

\subsection{Model limitations}

We propose in this section to discuss the limitation of the model developed in Section \ref{sec:model}. This latter is based on a basic assumption of the fluid velocity profile in-between pillars. This velocity profile is clearly non-physical because it does not respect the non-slipping conditions of the fluid velocity along walls. In order to solve exactly the problem of the fluid flow inside a confined micropillars array, the Stokes equation has to be solved with the corresponding boundary conditions and thus the resulting viscous force can be deduced. We expect our model to be close from the exact solution in the case of low pillar densities but to be less accurate for large pillar densities where the confinement modifies the fluid flow far from the assumed profile. Such an effect may explain the discrepancy between experiments and theoretical predictions observed for large pillar densities in Fig. \ref{fig:permeability_ratio}b.

Also, we assumed in Section \ref{sec:model_pillars} that the fluid flow adopts everywhere a fully developed velocity profile. The time needed for the fluid to reach such a state can be estimated. When a non-slipping boundary condition is imposed to a fluid flow, it spreads in a lateral direction $z$ according the relation: $z=\sqrt{\nu t}$ where $\nu=\eta/\rho$ is the kinematic viscosity of the liquid. In the case of the microchannels studied previously, the non-slipping conditions on the top and bottom walls spread in the $z$ direction according $z=\sqrt{\nu/D_0} \, x$. The fully developed velocity profile is reached when $z=h/2$ which occurs for a critical traveled distance $x_{min}/h = \sqrt{D_0/\nu}/2$. For $\eta=100 \, \rm{mPa \cdot s}$, $\rho=980 \, \rm{kg/m^3}$, $\gamma = 23 \, \rm{mN/m}$ and $h=80 \, \rm{\mu m}$ we estimate that $x_{min}/h \sim 10^{-3}$. As the microchannel is much longer than its height in our experiments, the assumption of a fully developed velocity profile is justified. 

\subsection{Comparison between open and confined micropillar arrays}

The presence of a top wall modifies the liquid impregnation behavior compared to the case of an open micropillar array. Such a wall introduces two counteracting ingredients on the capillary flow: an additional surface to wet and another no-slip boundary condition. Following the same procedure as in Section \ref{sec:model_pillars} with a free boundary condition at the top, we deduce a theoretical prediction for the diffusivity $D_{\rm{open}}$ in the case of an open microchannel. The ratio between the diffusivity in an open micropillar array and the diffusivity in a close one reads

\begin{equation}
\frac{D_{\rm{open}}}{D} = \left( \frac{ 1+ \left( 4 \overline{h} -1 \right) \phi }{1+ \left( 2 \overline{h} -1 \right) \phi } \right) \left[ \frac{ 1 + {\overline{h}^2}/{\sqrt{\frac{\phi_m}{\phi}} \, \left(\sqrt{\frac{\phi_m}{\phi}}-1 \right)^2}  }{  1 + {2 \overline{h}^2}/{\sqrt{\frac{\phi_m}{\phi}} \, \left(\sqrt{\frac{\phi_m}{\phi}}-1 \right)^2}  } \right]
\label{eq:ratio_open}
\end{equation}

\begin{figure}[htp!]
\centering
\includegraphics[width=8cm]{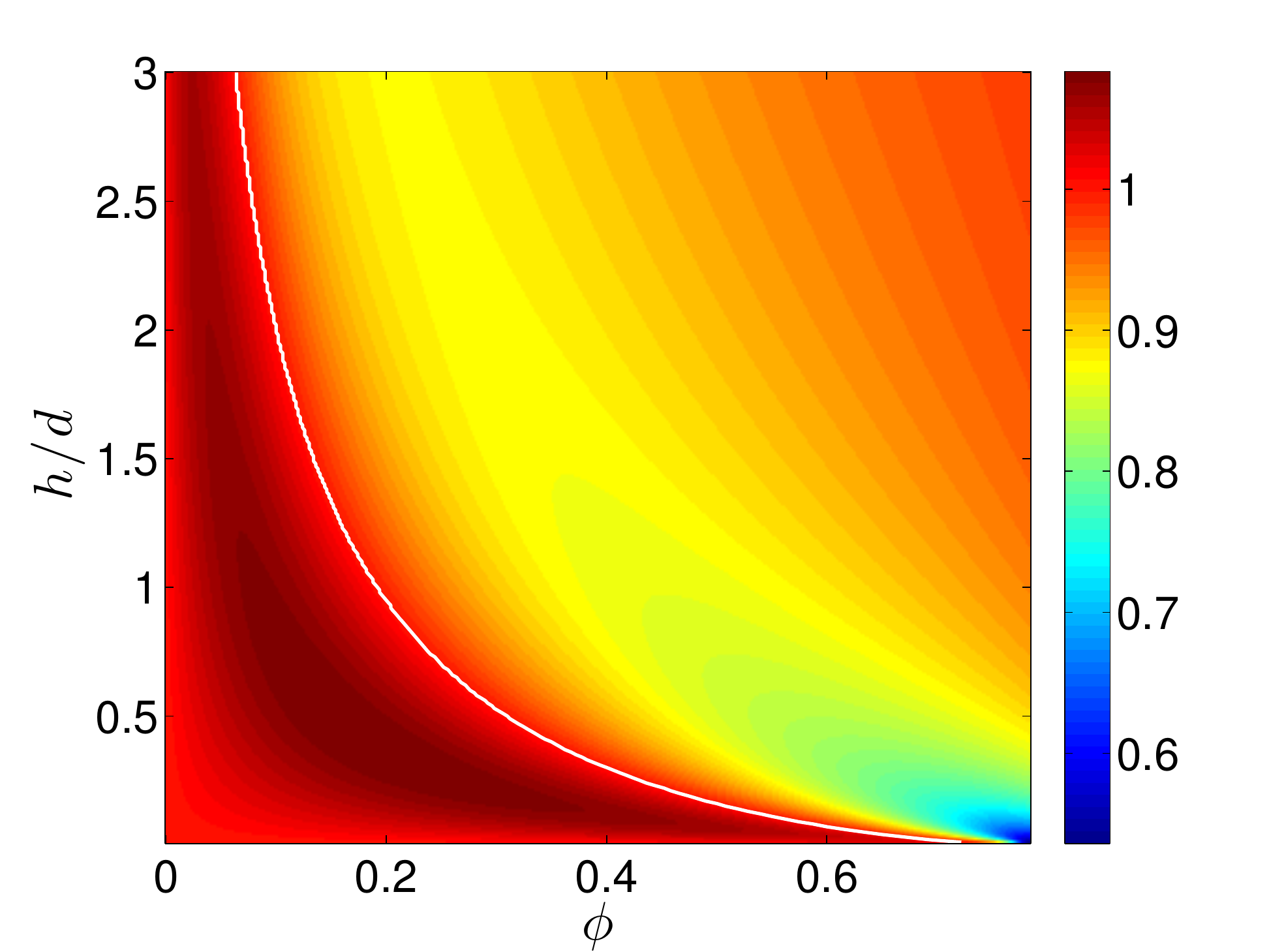}
\caption{Ratio $D_{\rm{open}}/D$ as a function of the pillar density $\phi$ and the normalized height of pillars $\overline{h}=h/d$ as predicted by equation (\ref{eq:ratio_open}). The white solid line separates the region where the diffusivity ratio $D_{\rm{open}}/D$ is larger than one.}
\label{fig:ratio_open}
\end{figure}

Figure \ref{fig:ratio_open} shows $D_{\rm{open}}/D$ as a function the pillar density $\phi$ and the normalized height $\overline{h}$ as predicted by Eq. (\ref{eq:ratio_open}). One notices that depending on the pillar aspect ratio and the pillar density, the diffusivity ratio $D_{\rm{open}}/D$ is lower or larger than one. Thus, the two antagonistic effects, \textit{i.e.} capillary pumping and viscous resistance, caused by the introduction of a top wall can overcome each other regarding the properties of the micropillar array. For dense arrays of slender pillars, the liquid impregnation in a confined micropillar array is faster than in a similar open array. In such situations, the confinement enhances the liquid spreading.

\section*{Conclusions}

The wicking of a micropillar array confined between two plates by a perfectly wetting and viscous liquid has been investigated experimentally. We showed that the presence of pillars can either enhance or slow down the dynamics of the contact line. However, the pillars only slow down the flow rate of the liquid which penetrates into the microchannel. A model based on the estimation of capillary forces and viscous friction inside the microchannel was developed. This latter provides a fair agreement with experimental data. In the end, this study predicts the fall of the efficiency of the liquid impregnation with increasing pillar density. This result has implications for the underfill process in microelectronics packaging where the liquid filling time is a limiting factor of the production (\cite{wan2008experimental}). 

A perspective of this work is to refine the basic model proposed to describe the liquid dynamics. The exact determination of the viscous force in the microchannel can be addressed by the way of numerical simulations. Besides, this study only considers pillars with cylindrical and square section. It could be interesting to vary this shape and introduce concave geometries in order to study the effect on the liquid dynamics and the possible entrapment of air bubbles. Furthermore, this study only considers the ideal situation where the liquid perfectly wets the entire surface of the microchannel. The modification of our predictions in the case of a liquid which partially wets the solid is an open question which has to be investigated. Finally, the impact of the non-Newtonian behaviors of a liquid on its wicking dynamics through a micropillar array can also be the scoop of future studies.

\begin{acknowledgements}

This research has been funded by the Inter-university Attraction Poles Programme (IAP 7/38 MicroMAST) initiated by the Belgian Science Policy Office. SD thanks the FNRS for financial support. We acknowledge Mathilde Reyssat, Tristan Gilet and Pierre Colinet for fruitful discussions and valuable comments. We are grateful to St\'ephanie Van Loo for precious advices concerning the realization of the PDMS microchannels.
\end{acknowledgements}

\section*{Appendix A}

Equation (\ref{eq:ratio}) allows to determine the critical pillar density $\phi_c$ in order to have $D=D_0$ with $\phi_c \neq 0$. The critical density verifies the relation

\begin{equation}
\phi_c \, \sqrt{\frac{\phi_m}{\phi_c}} \, \left(\sqrt{\frac{\phi_m}{\phi_c}}-1 \right)^2  = \frac{ \overline{h}^2}{2 \overline{h} - 1}
\label{eq:critical_density}
\end{equation}

\noindent which has a solution only if $\overline{h}>0.5$. Equation (\ref{eq:critical_density}) is solved numerically as a function of $\overline{h}$ and the solution is indicated in Fig. \ref{fig:ratio} by a white solid line.\\

For $\overline{h}>0.5$, the diffusivity ratio $D/D_0$ reaches a maximal value for a pillar density $\phi_{\rm{max}}$ which can be calculated by deriving Eq. (\ref{eq:ratio}) relatively to $\phi$ keeping $\overline{h}$ constant. Such a calculation is performed numerically and the solution is indicated in Fig. \ref{fig:ratio} by a white dashed line. Finally, the pillar aspect ratio $\overline{h}_{\rm{max}}$ which maximizes the diffusivity ratio $D/D_0$ for a given pillar density verifies

\begin{equation}
\begin{array}{l}
\phi \left( \sqrt{\frac{\phi_m}{\phi}} \, \left(\sqrt{\frac{\phi_m}{\phi}}-1 \right)^2  + \overline{h}_{\rm{max}}^2 \right) = \\
 \overline{h}_{\rm{max}}  \left( 1+ ( 2 \overline{h}_{\rm{max}} - 1) \phi \right)
 \end{array}
\label{eq:maximal_aspect_ratio}
\end{equation}

The solution of Eq. (\ref{eq:maximal_aspect_ratio}) is presented by a white dotted line in Fig. \ref{fig:ratio}.

\end{document}